\documentclass{epl}
\title{Statistical properties of driven Magnetohydrodynamic turbulence in three
dimensions: Novel universality}
\shorttitle{Statistical properties of magnetohydrodynamic turbulence}
\author{Abhik Basu \thanks{email:{basu@hmi.de}}}
\institute{Abteilung Theorie, Hahn-Meitner-Institut, Glienicker Strasse
  100, D-14109 Berlin, Germany, and
Poornaprajna Institute of Scientific Research, Bangalore, India.} 

\pacs{64.60Ht}{}
\pacs{ 0.5.70.Ln}{}
\pacs{47.65+a}{}
\begin{document}
\maketitle

\begin{abstract}
We analyse the universal properties of nonequilibrium steady states of driven
Magnetohydrodynamic (MHD) turbulence in three dimensions ($3d$). 
We elucidate the dependence of
various phenomenologically important dimensionless constants
on the symmetries of the two-point correlation functions. We, for the
first time, also
 suggest the intriguing possibility of multiscaling universality class 
varying continuously with certain dimensionless parameters. The experimental
and theoretical implications of our results are discussed.
\end{abstract}

In the vicinity of a critical point, equilibrium systems show universal scaling
properties for thermodynamic functions and correlation functions: These are
characterised by universal scaling exponents that depend on the spatial
dimension $d$ and the symmetry of the order parameter (e.g., Ising, XY etc.)
\cite{fisherrev}, but not on the parameters that specify the details of the
Hamiltonian. A notable exception is the class of two-dimensional models, such as the XY
model in $d=2$, in which continuously varying scaling exponents are found \cite{chai}.
Dynamic scaling exponents, that characterise the behaviour of time-dependent correlation
functions also show universality \cite{hal}. A different situation arises in 
driven, dissipative, nonequilibrium
systems with nonequilibrium statistical steady states (NESS).
We show here that driven homogeneous and isotropic 
MHD turbulence in three dimensions ($3d$) 
is a good natural candidate
for a system with an NESS whose universal properties 
vary continuously with the degree of crosscorrelations between the velocity
and magnetic fields.  
Our results illustrate the sensitivities of the
statistical properties of nonequilibrium steady states on the parameters of the
model, a situation which does not arise in equilibrium systems.

Turbulence in plasmas is described by the equations of MHD
\cite{jack} in $3d$ for the coupled evolution of the
velocity field $\bf v$ and the magnetic field $\bf b$. Solar-wind data 
\cite{grau} and numerical simulations of the $3d$MHD equations as well as shell
models \cite{abprl} show multiscaling corrections to the simple Kolmogorov (K41) 
scaling
\cite{kol} for the structure functions of $\bf v$ and $\bf b$: $S_a^p(r) \equiv
\langle [|a_i({\bf x+r})-a_i({\bf x})|^p]\rangle \sim r^{\zeta_p^a},\,a=v,b$ 
for $r$ in the inertial range
between the forcing scale $L$ and the dissipation scale $\eta_d$ (i.e., $L\gg r\gg
\eta_d$). 
In observations and
experiments magnetic Prandtl number $P_m$(=the ratio of the magnetic to kinetic viscosity)
and $e\equiv {E_m\over E_v}$ (the ratio of the magnetic- to the kinetic- energy)
are known to have a wide range of values: $e\ll
1\Rightarrow$ kinetic regime of $3d$MHD (as on the solar surface),
$e\sim O(1)\Rightarrow$ equipartition regime (e.g., solar wind in its rest
frame) and $e\gg 1\Rightarrow$ magnetic regime (e.g., fusion plasmas)
\cite{mont}. Similarly $P_m$ can be high ($>
1$), e.g., in galactic and proto galactic plasmas \cite{astro}. 
Low values of $P_m$ 
are found in liquid metals \cite{liq}.  In numerical simulations, the
structure functions of $\bf v$ and $\bf b$ in $3d$ exhibit multiscaling
with exponents different from their fluid analogues \cite{abprl,cho}. 
In
particular Ref.\cite{abprl} elucidated different multiscaling exponents for $\bf
v$ and $\bf b$ (i.e., $\zeta_p^v\neq\zeta_p^b$).

In this letter we develop a one-loop, self-consistent theory for the randomly forced
$3d$MHD equations and use it to calculate several interesting results concerning
 its NESS. We show, in
particular, that (i) the variation of $e$ and renormalised (effective) $P_m$ are 
connected  to each other
(ii) $P_m,\,e$, Kolmogorov's constants $K_v$ and $K_b$ (see below) 
vary with the degree of crosscorrelations between
$\bf v$ and $\bf b$. 
We calculate the intermittency exponents (these in a log-normal model 
provide approximate
estimates of the deviations from the K41 scaling) which also vary
continuously with the degree of crosscorrelations between $\bf v$ and $\bf b$.
This suggests
an interesting possibility of continuously varying multiscaling 
universality classes for
$3d$MHD. We 
show that these variations are linked with the ratios of the different 
 two-point correlation functions. 
Our observations on multiscaling
of $3d$MHD based on our self-consistent calculations on a log-normal model allows
for different multiscaling exponents for the velocity and magnetic fields as
reported in Ref. \cite{abprl}.

The $3d$MHD equations consist of the Navier-Stokes equation for the velocity
field $\bf v$ supplemented by the Lorentz force due to the magnetic field $\bf
b$:
\begin{equation}
{\partial {\bf v}\over \partial t}+ {\bf
v}.\nabla {\bf v}=-{\nabla p\over \rho} +{(\nabla
\times {\bf b})\times {\bf b}\over\ 4\pi \rho} +\nu_o \nabla^2 {\bf v}
+{\bf f}
\label{mhdu}
\end{equation}
with $\nabla {\bf .v}=0$ (incompressibility) and 
Amp\`ere's law for a conducting fluid 
\begin{equation}
{\partial {\bf b}\over\partial t}+{\bf v.\nabla
b}={\bf b.\nabla v} +\mu_o \nabla^2 {\bf b} +
{\bf g}.
\label{mhdb}
\end{equation}
In (\ref{mhdu}) and (\ref{mhdb}), $\rho$ is the fluid
density, $p$ is the pressure, $\mu_o$ is the ``magnetic
viscosity'', arising from the nonzero resistivity of the plasma, $\nu_o$ is the
kinematic viscosity and $\bf f$ and $\bf g$ are forcing functions. 
Under rescaling of space and time  
 ${\bf v}({\bf x},t)$ and ${\bf b}({\bf x},t)$ scale as 
$v_i(l{\bf x},l^zt)\rightarrow l^{\chi_1}v_i,\,b_i(l{\bf x},b^zt)
\rightarrow l^{\chi_2}b_i({\bf x},t)$. 
Since $\bf v$ and $\bf b$ are polar and axial vectors respectively, 
crosscorrelation tensor $\langle v_i({\bf k},t)b_j({\bf -k},0)\rangle$ is odd
and imaginary in wavevector $\bf k$ \cite{abjkb}. 

We employ a self-consistent mode coupling scheme (SCMC).
The response and the correlation functions of the fields $\bf
v$ and $\bf b$ are defined by $v_i\equiv G_v f_i,\, b_i\equiv G_b g_i,\,$ and
correlators $C_{ij}^v
\equiv \langle v_i({\bf k},t)v_j({\bf -k},0)\rangle,\,C_{ij}^b\equiv \langle
b_i({\bf k},t)b_j({\bf -k},0)\rangle$ with the crosscorrelator $C_{ij}^c \equiv
\langle v_i({\bf k},t)b_j({\bf -k},0)\rangle$. In the scaling limit, 
in terms of the dynamic exponent $z$ and the roughness exponents 
$\chi_1$ and $\chi_2$,
$G_v^{-1}=i\omega -\Sigma_v (k,\omega),\,G_b^{-1}=i\omega -\Sigma_b (k,\omega)$
where $\Sigma_v=k^{-z}\eta_v(\omega/k^z),\, \Sigma_b=k^{-z}\eta_b(\omega/k^z)$
are the self-energies whereas the correlators are given by
$C_{ij}^v=P_{ij}k^{-d-2\chi_1-z}\eta_1 (\omega/k^z),\, C_{ij}^b= P_{ij}
k^{-d-2\chi_2-z}\eta_2 (\omega/k^z)$. $\eta_v(\omega/k^z), \eta_b(\omega/k^z),
\eta_1 (\omega/k^z), \eta_2 (\omega/k^z)$ are scaling functions and $P_{ij}$ is
the transverse projection operator. In systems out of equilibrium, there is no
particular relation between the noise variance and the dissipation coefficient,
unlike their equilibrium counterparts where such a relation exists due to the
Fluctuation-Dissipation-Theorem. It is also well-known that the statistical
properties of the nonequilibrium systems depend strongly on the noise variances.
We, here in particular, take the forcing terms $\bf f,\,g$ to be 
Gaussian noises with zero mean and 
covariances proportional to $k^{-d}$ in $d$-dimension
\cite{yakhot}.
This ensures that, in absence of a mean magnetic field, the energy spectra are
K41-like \cite{abprl,kol}.
In this model, the advective nonlinearities and the noise variances do not
renormalize. This means under simultaneous rescaling of space and time,
the nonlinear vertices and the noise variances are affected only by their naive
scaling dimensions. At the fixed point, 
these conditions leads to {\em strong dynamical scaling}, i.e., same dynamic
exponent for both the fields and immediately yield
 $z=2/3,\,\chi_1=\chi_2=\chi=1/3$ corresponding to
K41 scaling \cite{abjkb,others}. Hence the ratios of various correlations are 
dimensionless numbers. 
We assume Lorentzian line
shapes for the self-energies and the correlation functions. The fact that the
values of the scaling exponents $z$ and $\chi$ are dimension independent
is because the noise correlations change as a function
of dimensionality in such a way as to render the exponents dimension independent.

We start with the
zero-frequency forms for the self-energies (or the relaxation rates) 
$\Sigma_v(k,\omega=0)=\nu k^z,\,\,\Sigma_b(k,\omega=0)=\mu k^z,$
and the correlations 
$C_{ij}^v(k,\omega=0)={2D_1\over\nu^2}P_{ij}
k^{-2\chi-d-z},\,$
$C_{ij}^b(k,\omega=0)={2D_2\over\mu^2}P_{ij}k^{-2\chi-z-d},$
in $d$-dimension.
The crosscorrelation tensor $C_{ij}^c$ in general has both symmetric $C_{ij}^s$ 
and antisymmetric $C_{ij}^a$ parts: $C_{ij}^c=C_{ij}^s+C_{ij}^a;$
$C_{ij}^s(k,\omega=0)= {2i\tilde{D}({\bf k})P_{ij}\over\nu\mu}k^{-2\chi-z-d},$
in $d$-dimension and with $\tilde {D}({\bf k})=-
\tilde{D}({\bf -k});\, \tilde{D}({\bf k})\tilde{D}({\bf k})=(\tilde{D})^2$
  and 
$C_{ij}^a(k,\omega=0)={2i\hat{D}\over\nu\mu}\epsilon_{ijp}k_pk^{-2\chi-z-4}$
(this form is for $3d$ only). 
The mean crosshelicity $\sum_{\bf k}\langle \bf v(k).b(-k)\rangle$ and 
mean electromotive force $\sum_{\bf k}\langle \bf v(k) \times b(-k)\rangle$ are
proportional to $\tilde D$ and $\hat D$ respectively.
Here $\epsilon_{ijp}$ is the totally
antisymmetric tensor in $3d$.
Previous renormalisation group/self-consistent
approach to $3d$MHD did not take any crosscorrelation function \cite{others}
fully into consideration. The noises in stochstically driven models, such as Eqs.
(\ref{mhdu}) and (\ref{mhdb}), represent coarse-grained
 effects of the degrees of freedom
whose time-scales of decay are faster than the coarse-grained variables ($\bf 
v$ and
$\bf 
b$) used. Since both the noises in Eqs. (\ref{mhdu}) and (\ref{mhdb}) have same
microscopic origin there is no apriori reason for their crosscorrelations to
vanish always. 
In SCMC approach vertex corrections are
neglected. Lack of vertex renormalisations in the zero wavevector limit 
in $3d$MHD allows SCMC to yield {\em exact} relations between the scaling
exponents $z$ and $\chi$, as in the
the noisy Burgers/Kardar-Parisi-Zhang equation \cite{fns}.
In the context of the noisy Burgers equation in 1+1 dimensions, Frey
{\em et al} have shown, by using nonrenormalisation of the advective
nonlinearities and a second-order perturbation theory, that the effects of the
vertex corrections on the correlation functions are very small \cite{hwa}. Similar
conclusions for $3d$MHD will presumably follow from the vertex nonrenormalisations
in $3d$MHD, though a rigorous calculation is still lacking.

One may note that the symmetries of the symmetric and the antisymmetric parts of
the crosscorelations are different under the exchange of the cartesian indices. 
Hence, the presence of one does not lead to the generation of the other by
non-linear interactions. Thus it is possible to treat them separately without any
loss of generality.
It is straight forward to extend  the various expressions 
derived in the paper for the case when both the symmetric and the
anti-symmetric parts are present.

{\em Symmetric Cross-correlations.-}
We obtain the one-loop self-consistent integral equations
 for the self-energies and correlation
functions. The one-loop diagrammatic corrections to
the crosscorrelation function vanish as it is odd in momentum (see, e.g., 
\cite{decay}). We first
consider the case when $\hat{D}=0$, i.e., no antisymmetric part of the
crosscorrelations. By matching at $\omega=0$,
 for the self-energies we obtain (by using $\chi=1/3,\,z=2/3$) in any
dimension $d$
\begin{equation}
\nu={3D_1\over 4\nu^2}{S_d\over (2\pi)^d}{d-1\over (d+2)}+{3D_2\over 4\mu^2}
{S_d\over (2\pi)^d}{d^2+d-4\over d(d+2)},
\label{selfnu}
\end{equation}
\begin{equation}
\mu={3D_1\over 2\nu(\nu+\mu)}{S_d\over (2\pi)^d}(1 - {1\over d})+ {3D_2\over 2\mu 
(\nu +\mu)}{S_d\over (2\pi)^d}(1 - {3\over d}).
\label{selfmu}
\end{equation}
Similarly, by demanding consistency in the amplitudes of the correlations we
get
\begin{equation}
{D_1\over D_2}={\left[{D_1^2\over\nu^3}+{D_2^2\over\mu^3}-{4\tilde{D}^2\over\nu
\mu (\nu+\mu)}\right]\left[1-{2\over d}+{2\over d(d+2)}\right]\over \left[{4
D_1D_2\over\nu\mu(\nu+\mu)^3}+{16\tilde{D}^2\over (\nu+\mu)^3}\right](1-{2\over
d})}.
\label{selfcor}
\end{equation}

In terms of the {\em renormalised magnetic Prandtl number} $P_m=\mu/\nu$ and 
$\Gamma=D_2/D_1$ we can write
 Eqs.(\ref{selfnu}), (\ref{selfmu}) and (\ref{selfcor}) as
\begin{equation}
P_m^{-1}={{d-1\over 2(d+2)}+{\Gamma\over 2P_m^2}\left[ {d^2+d-4\over
d(d+2)}\right]\over {1\over 1+P_m}\left(1-{1\over d}\right) +{\Gamma\over P_m(1+P_m)
} \left(1-{3\over d}\right)},
\label{self1}
\end{equation}
\begin{equation}
{\rm and}\; \Gamma={\left[1+{\Gamma^2\over P_m^3}-{4\tilde{\beta}
\over P_m(1+P_m)}\right] \left[
1-{2\over d}+{2\over d(d+2)}\right]\over \left[{4\Gamma^{-1}\over P_m(1+P_m)}+ {16
\tilde{\beta}\over (1+P_m)^3\Gamma^2} \right](1-{2\over d})},
\label{self2}
\end{equation}
where $\tilde{\beta}=({\tilde{D}\over D_1})^2$. Equations (\ref{self1}) and
(\ref{self2}) {\em explicitly} demonstrate that 
$P_m$ and the magnetic- to kinetic- energy ratio $e={\Gamma\over P_m}$ depend on
each other.
When $\tilde{\beta}=0$,
i.e., zero crosshelicity,  Eqs.(\ref{self1}) and (\ref{self2})
yield $P_m=0.67$ and $\Gamma=0.63$  and hence energy-ratio 
$e=\Gamma/P_m\approx 0.94$ in $3d$. Thus for zero
crosshelicity $P_m$ and $e$ are {\em fixed} numbers. But with finite
crosshelicity, i.e., a finite $\tilde{\beta}$ in Eqs.(\ref{self1}) and
(\ref{self2}), $P_m$ and $\Gamma$ and hence $e$ depend on $\tilde{\beta}$ in
any dimension $d$. For $\tilde\beta>0$ few representative values of $P_m$
and $\Gamma$ in $3d$ are given in Table 1.

{\em Anti-Symmetric Cross-Correlations.--}
For the effects of the antisymmetric part 
(under the exchange of  $i$ and $j,\, i,j$ are
Cartesian indices) of the crosscorrelation function
in $3d$ on the long wavelength properties
we again use self-consistent methods. 
Equations (\ref{selfnu}) and (\ref{selfmu}) remain unaltered. However,
Eq.(\ref{selfcor}) is changed to
\begin{equation}
{D_1\over D_2}={\left[{D_1^2\over\nu^3}+{D_2^2\over\mu^3}\right]{2\over 5}+
{4\hat{D}^2\over 3\nu\mu (\nu+\mu)}\over \left[{4 D_1D_2\over 
3\nu\mu(\nu+\mu)} -{16\hat{D}^2\over 3(\nu+\mu)^3}\right]}.
\label{selfcor1}
\end{equation}
Again by defining 
$\hat{\beta}=({\hat{D}\over
D_1})^2$, we obtain
self-consistently
\begin{equation}
\Gamma={\left[1+{\Gamma^2\over P_m^3}\right]
{2\over 5}+{4\hat{\beta}\over 3(1+P_m)}
\over  \left[{4\Gamma^{-1}\over 3P_m(1+P_m)}-{16\hat{\beta}\over 3
(1+P_m)^3}\right]}.
\label{self3}
\end{equation}
Thus Eqs.(\ref{self1}) and (\ref{self3}) together provide self-consistent
relations between $P_m$ and $\Gamma$ and hence between $P_m$ and $e$ when
there is a finite antisymmetric crosscorrelation. 
A few representative values of
$P_m$ and $\Gamma$ for $\hat\beta >0$ are given in Table 1.

{\em Kolmogorov's constants.--}
According to the Kolmogorov's hypothesis for fluid turbulence \cite{kol} in the
inertial range energy spectrum 
$E(k)=K_o\epsilon^{2/3}k^{-5/3}$, where $K_o$, a {\em universal}
constant, is the Kolmogorov's constant and $\epsilon$ is the 
energy dissipation rate
per unit mass. Various calculations, based on different techniques by different
groups \cite{krai,leslie,yakhot,jkb} show that $K_o\sim 1.5$ in three
dimensions. 
For MHD, by using Novikov's theorem \cite{nov}, we connect the 
dissipation of the total energy with
the {\em sum} of the velocity and magnetic field correlation-amplitudes. 
We can then relate the dissipation of the 
{\em reduced energy} $E_R=E_v-E_b$ with the
correlation-amplitudes.  Thus,
in the lowest order of the perturbation theory we
calculate the coefficients of the velocity and magnetic field
correlations in terms of dissipation rates: 
$E_v(k)=K_v \epsilon_v
^{2/3}k^{-5/3};\;E_b(k)=K_b \epsilon_b ^{2/3}k^{-5/3}$. $K_v$ and $K_b$ are the
Kolmogorov's constants for MHD, and $\epsilon_v$ and $\epsilon_b$ are the mean 
dissipation rates of the kinetic energy and magnetic energy respectively. 
In absence of a mean magnetic field energy spectra scale
as $k^{-5/3}$ in the inertial range \cite{abprl}. In Ref.\cite{verma} 
Kolmogorov's
constants for MHD turbulence (for the  Els\"asser variables) have been 
calculated incorrectly
for various energy-ratio $e$ due to the absence of crosscorrelation functions 
of appropriate
structure. We
work out the Kolmogorov's constants with finite crosscorreletion functions. 
From Eq.(\ref{self1}) in dimension $d=3$, we have
 $\nu=[{0.6D_1\over 4}{S_3\over (2\pi)^3}]^{1\over 3}(1+{4\over
3}{\Gamma\over P_m^2})^{1\over 3}$.
In our notations, $\langle v_i({\bf k},\omega)v_j({\bf -k},-\omega)\rangle = 
{2D_1P_{ij}k^{-3}\over \omega^2+\nu^2k^{2z}}$ which in $3d$ gives (in the
inertial range) for kinetic energy spectrum 
$E_v(k)
=1.186 [2D_1(1+.62\{1+{\Gamma^2\over P_m^3}-{4a\over
P_m(1+P_m)}\})]^{2\over 3}{S_3\over (2\pi)^3}
{k^{-5/3}\over (1+1.33{\Gamma\over P_m^2})^{1/3}}.$
To connect dissipation rate
$\epsilon_v$ with noise correlation amplitude $D_1$ we use Novikov's theorem
\cite{nov}
which gives $2D_1{S_3\over (2\pi)^3}=\epsilon_v$. Substituting this in $E_v(k)$
above, we 
 obtain Kolmogorov's constant
\begin{equation}
K_v=1.186{[1+.62\{1+{\Gamma^2\over P_m^3}-{4a\over
P_m(1+P_m)}\}]^{2/3}\over (1+1.33 {\Gamma\over P_m^2})^{1/3}},
\label{kolv}
\end{equation}
where $a=\tilde{\beta},\,-\hat{\beta}$ for finite symmetric and antisymmetric
crosscorrelations respectively. A similar calculation yields $K_b=K_v$.
Our results show, unsurprisingly, that 
$K_v$ and $K_b$ are {\em not fixed} numbers, rather they
depend upon other dimensionless numbers like magnetic Prandtl number $P_m$
and energy-ratio $e$. For $\Gamma=0$, i.e., for no magnetic fields
we get the result $K_v=1.63$ for pure fluid turbulence
which is well-within the range of accepted values \cite{leslie,belini}.

{\em Possibilities of variable multifractality.---}
Experiments and numerical sumulations \cite{abprl,cho} find nonlinear multiscaling
corrections to the K41 prediction of $\zeta_p^a=p/3$ for the structure functions
in the inertial range. Until the date, no controlled perturbative calculation for
$\zeta_p^a$ is available.
To account for multiscaling in fluid turbulence, however,
Obukhov \cite{obu} and Kolmogorov \cite{kol2} assumed a log-normal distribution
for dissipation $\epsilon$ to arrive at
$S_p^v(r)=
\langle |\Delta v|^p\rangle=C_p\overline{\epsilon}^{p/3}r^{p/3}\left({L\over r}
\right)^{(\overline{\delta}/2)p(p-3)},$
where $\overline{\epsilon}$ is the mean value of $\epsilon$ and
$\langle
\epsilon ({\bf x+r})\epsilon ({\bf x})\rangle\propto \langle (\Delta
v)^6/r^2\rangle\sim (L/r)^{9\overline{\delta}}. $
 For small $\overline\delta$,
$\delta\simeq 9\overline{\delta}$. A standard calculation on the randomly
stirred model yields {\em intermittency exponent} $\delta=0.2$ \cite{amita} where
$\delta=9\overline\delta$.
whereas the best possible estimates from experiments is 0.23 \cite{amita}.
In MHD dissipations of kinetic- ($\epsilon_v$) and magnetic- ($\epsilon_b$)
energies fluctuate in space and time. Consequently we 
define two intermittency exponents $\delta_v,\,\delta_b$ 
for the kinetic [$\epsilon_v={1\over 2}\nu
\sum_{\alpha\beta}\left({\partial v_{\alpha}\over\partial
x_{\beta}}+{\partial v_{\beta}\over\partial x_{\alpha}}\right)^2$],
and magnetic energy dissipations [$\epsilon_b={1\over 2}\mu
\sum_{\alpha\beta}\left({\partial
b_{\alpha}\over\partial
x_{\beta}}+{\partial b_{\beta}\over\partial x_{\alpha}}\right)^2\,$]
to explain the
deviations from the K41 scaling. 
We calculate the exponents below by using the
Eqs.(\ref{mhdu}) and (\ref{mhdb}) following closely Ref.\cite{amita}.
%
We work with the self-consistent forms for the
self-energies and correlation functions given above along with the
consistency relations for the amplitude-ratios $\Gamma,\tilde{\beta},
\hat{\beta}$ and $P_m$. 

Following Ref.\cite{amita}, we find the dissipation correlation functions in $3d$
to be 
$\langle\epsilon_v({\bf x+r})\epsilon_v({\bf x})\rangle\simeq
12.4\epsilon_v^{2}\alpha_v^2 K_v^2 \ln {L\over r}, \,
\langle\epsilon_b({\bf x+r})\epsilon_b({\bf x})\rangle
\simeq 12.4\epsilon_v^{2}\alpha_v^2 \Gamma^2K_v^2 \ln {L\over r}.$
$\alpha_v$ is defined by the relation $\nu=\alpha_v\epsilon_v^{1/3}$. By using
Eqs.(\ref{selfnu}), (\ref{selfmu}), (\ref{self1}), (\ref{self2}) and
(\ref{self3}) we find,  in $3d$,
\begin{eqnarray}
\alpha_v=0.4[1.62+0.62({\Gamma^2\over P_m^3}-{4a\over
P_m(1+P_m)})]^{1/3}[1+1.33{\Gamma\over P_m^2}]^{1/3}, 
\label{alphaeq}
\end{eqnarray}
where $a=\tilde{\beta}$ and $-\hat{\beta}$
for finite symmetric and antisymmetric crosscorrelations respectively.
For the pure fluid case $\alpha_v\simeq 0.5$ \cite{yakhot} and is {\em
universal}. For MHD, however we see that $\alpha_v$ varies with $\tilde{\beta}$
and $\hat{\beta}$, similar to $K_v$. 
We then obtain fluid intermittency exponent
\begin{eqnarray}
\delta_v&=&0.2\left[1+0.4\left({\Gamma^2\over P_m^3}-{4a\over P_m(1+P_m)}
\right)\right]^{4/3},
\label{delv}
\end{eqnarray}
where, $a$ is the same as before.
A similar calculation obtains magnetic intermittency exponent 
$\delta_b=\Gamma^2\delta_v$. As expected, both 
$\delta_v$ and $\delta_b$ vary with $P_m$ and $\Gamma$ (or $e$). In general, the
intermittency corrections to the simple K41 scaling of the $\bf v$ and $\bf b$
structure functions are {\em unequal}, as has been seen in the direct numerical
simulations (DNS) and shell-model studies  \cite{abprl}.
The presence of multiple intermittency exponents makes it difficult to 
directly extend the fluid
log-normal intermittency model to $3d$MHD. However, the continuous variations of the
intermittency exponents with $\tilde{\beta}$ and $\hat{\beta}$ strongly suggest 
{\em continuously varying multiscaling universality classes}
 for $3d$MHD. 

In our calculations, $\tilde{\beta}$ and $\hat\beta$ parametrise the internal
symmetries of the system. They represent
the relative strengths of the symmetric and anti-symmetric parts of the
crosscorrelation function. Therefore, $\tilde{\beta}$ and $\hat\beta$ charactrerise
the symmetries of the two-point correlation functions under inversion of the
parity and exchange of the cartesian indices.
Results from our log-normal 
model-type calculations open up an intriguing possibility of multiscaling
properties varying continuously with symmetries parametrised by the
dimensionless numbers described above. 
We believe that our results (despite the limitations of mode coupling theories -
see, e.g., \cite{yakhot,eyink}) will
provide a valuable insight to the understanding of the difference 
in the multiscaling
properties of the velocity and magnetic fields \cite{abprl} and may also explain
the somewhat different numerical estimates of the multiscaling exponents by
different groups \cite{abprl,cho}.
We believe our results will stimulate further numerical as well as experimental 
studies for the measurements of the dimensionless parameters introduced above.
To check our results we suggest
MHD experiments on liquid sodium systems with active grids and random passage of
current. Various $\tilde{\beta}$ and 
$\hat{\beta}$ can be achieved in experiments/simulations by controlling the
appropriate Grashof numbers \cite{gras}, constructed out of the various 
noise correlators in stochastically driven MHD and by measuring the cross
helicity, the electromotive force and the kinetic energy in the steady state. 
Renormalised $P_m$ can be
calculated by measuring time-dependent correlations.
In closing we note that many natural systems, e.g., solar wind, 
plasmas in tokamac have strong mean
magnetic fields making them anisotropic. Such effects
can also be included in our scheme of calculations by
appropriately modifying the form of the self-energies.

From a broader point of view, our results unveil qualitatively new properties in
the steady-states of nonequilibrium driven, difusive systems. In contrast to the
more well-known equilibrium systems, we illustrate the
richness of nonequilibrium phenomenologies by using $3d$MHD as an example. 
Recently, Drossel {\em et al}~\cite{barbara},
in a set of coupled Langevin equations describing the
interplay between phase ordering dynamics in the bulk and roughening
dynamics of the interface of binary films, find a similar
continuous variation of the dynamical exponent with the coupling
strength of the bulk and surface fields. Recently, 
continuously varying universal properties in the context
of a simple coupled Burgers-like model has been discussed in Ref.\cite{abfrey}. 
Here, for the first time, we
illustrate a natural realisation of continuously varying multiscaling universality
class in $3d$MHD.  We believe our
studies will provide further stimuli to the studies of these new aspects of driven
nonequilibrium systems.

The author thanks the Alexander von Humboldt Stiftung, Germany for financial
support and R. Pandit, E. Frey and J. Santos for their critical
comments on the manuscript.
\vskip-0.5cm

\begin{table}
\caption{Representative values of $P_m$, $\Gamma,\,e,\,K_v,\,\alpha_v$ and 
$\delta_v$ as functions of $\tilde{\beta}$ and $\hat{\beta}.$}
\label{table1}
\begin{tabular}{|c|c|c|c|c|c|c|c|}
$P_m$ & $\Gamma$ & $e$ & $\tilde\beta$ & $\hat\beta$ & $K_v$ & $\alpha_v$ &
$\delta_v$ \\ \hline
- & 0 & 0 & 0& 0&1.6 & 0.47 & 0.2 \\ \hline
1 & 1.76 & 1.76 & 3& 0& 0.6& 0.6& 0.04\\ \hline
0.1 & 0.06 & 0.6 & 0& 0.15&3.5 & 1.6& 10\\ \hline
0.67 & 0.63 & 0.94 & 0& 0 & 1.52& 0.76&0.6\\ \hline
\end{tabular}
\end{table}

\end{document}